\def\b{\bibitem}
\def\be{\begin{equation}}
\def\ee{\end{equation}}
\def\bea{\begin{eqnarray}}
\def\eea{\end{eqnarray}}
\def\bml{\begin{mathletters}}
\def\eml{\end{mathletters}}
\documentstyle[aps,prb,eqsecnum,psfig,epsf,floats]{revtex}
\draft
\begin{document}
% Macros for the various macro package names, etc.
\def\SNG{{\em Physical Review Style and Notation Guide}}
\def\LUG {{\em \LaTeX{} User's Guide \& Reference Manual}}
\def\btt#1{{\tt$\backslash$\string#1}}%
\def\REVTeX{REV\TeX}
\def\AmS{{\protect\the\textfont2
        A\kern-.1667em\lower.5ex\hbox{M}\kern-.125emS}}
\def\AmSLaTeX{\AmS-\LaTeX}
\def\BibTeX{\rm B{\sc ib}\TeX}
%\makeatletter
%\tighten
\twocolumn[\hsize\textwidth\columnwidth\hsize\csname@twocolumnfalse%
\endcsname
\title{Local versus Nonlocal Order Parameter Field Theories for Quantum Phase 
       Transitions%\\
       %\small{$[$Phys. Rev. B {\bf xx}, xxx (200x)$]$}
                                                   }
\author{D. Belitz}
\address{Department of Physics and Materials Science Institute,%\\
University of Oregon,%\\
Eugene, OR 97403}
\author{T.R. Kirkpatrick}
\address{Institute for Physical Science and Technology, 
         and Department of Physics\\
         University of Maryland, College Park, MD 20742}
\author{Thomas Vojta}
\address{Theoretical Physics, Oxford University, Oxford, UK\\
         and Department of Physics, University of Missouri-Rolla, Rolla, 
         MO 65409}
\date{\today}
\maketitle
 
\begin{abstract}
General conditions are formulated that allow to determine which quantum
phase transitions in itinerant electron systems 
can be described by a local Landau-Ginzburg-Wilson or LGW theory solely
in terms of the order parameter. A crucial question is the degree to
which the order parameter fluctuations couple to other soft modes.
Three general classes of zero-wavenumber order parameters, in the particle-hole
spin-singlet and spin-triplet channels, and in the particle-particle channel,
respectively, are considered. It is shown that the particle-hole spin-singlet
class does allow for a local LGW theory, while the other two classes do
not. The implications of this result for the critical behavior at various
quantum phase transitions are discussed, as is the connection with 
nonanalyticities in the wavenumber dependence of order parameter
susceptibilities in the disordered phase.
\end{abstract}

]
%\narrowtext
\section{Introduction}
\label{sec:I}

Much interest in the field of quantum many-body physics has recently
focused on quantum phase transitions.\cite{Hertz,Sondhi_et_al,Sachdev}
These are phase transitions
that occur at zero temperature as a function of some non-thermal control
parameter, often pressure or composition, and are driven by quantum
fluctuations as opposed to thermal ones. In addition to being of fundamental
interest, quantum phase transitions are important because they are
believed to underly a number of interesting low-temperature phenomena,
in particular various forms of exotic 
superconductivity.\cite{SB_Proceedings,Saxena_et_al}

Hertz, building on earlier work, has given a general scheme for the
theoretical treatment of quantum phase transitions.\cite{Hertz} 
After identifying
the order parameter of interest, one performs a Hubbard-Stratonovich
decoupling of the interaction term responsible for the ordering, with
the order parameter field as the Hubbard-Stratonovich field. One then
integrates out the fermions to obtain a field theory entirely in terms
of the order parameter. This Landau-Ginzburg-Wilson (LGW) theory can
then be analyzed by means of renormalization-group techniques. To the
extent that the LGW theory is well behaved, the resulting critical
behavior in three-dimensions ($3$-$d$) is often mean-field like, since 
the quantum phase transition is related to the corresponding classical 
transition in a higher dimension. Until recently, it therefore was
believed that most quantum phase transitions are not interesting from
a critical phenomena point of view.

In recent years it has become clear that in general there are problems
with Hertz's scheme. In particular, for one of the most obvious examples
of a quantum phase transition, viz. the zero-temperature transition in
itinerant ferromagnets, it was shown that Hertz's method does not lead 
to a local quantum field theory.\cite{us_fm_dirty,us_fm_clean} 
This is because there are soft modes other than
the order parameter fluctuations, specifically, soft particle-hole
excitations other than the spin-density fluctuations, that couple to the
order parameter. Since these ``additional'' soft modes are integrated
out in deriving the LGW functional, the resulting field theory has
vertices that are not finite in the limit of vanishing wavenumbers 
and frequencies. Such nonlocal field theories are hard to analyze, and
unsuitable for explicit calculations. A better, albeit more
involved, strategy in such cases is to integrate out the massive modes
only, keep all of the soft modes on equal footing, and derive a coupled
field theory for the latter. Such a procedure for the quantum ferromagnetic
transition in the presence of quenched disorder has recently revealed 
that the critical behavior is not
mean-field like as suggested by Hertz, and not even given by a simple
Gaussian fixed point as proposed on the basis of the nonlocal
LGW theory,\cite{us_fm_dirty} but rather given by the
power laws of the simple Gaussian fixed point with complicated multiplicative
logarithmic corrections to scaling.\cite{us_fm_local} In clean 
itinerant ferromagnets,
Hertz theory also breaks down\cite{us_fm_clean} and the quantum phase
transition is generically of first order.\cite{us_1st_order}

This example of a breakdown of the order parameter field theory method
casts serious doubt on the very concept of an LGW theory for quantum
phase transitions.\cite{thermal_footnote} However, an equally prominent
example for which the LGW concept works, at least in 
$3$-$d$,\cite{AbanovChubukov} is given by the quantum
antiferromagnetic transition, which was also discussed by Hertz.\cite{Hertz}
This raises the following question: Which quantum phase transitions
in itinerant electron systems can
be described by a local order parameter field theory, and which require a
more complicated analysis in terms of a coupled field theory? The quantum
ferromagnetic and antiferromagnetic transitions, respectively,
in $3$-$d$ provide examples for each of these two categories.

It is the purpose of the present paper to provide a partial answer to
this question. Specifically, for quantum phase transitions with zero-wavenumber
order parameters we will provide a classification scheme that
shows for which transitions Hertz's scheme works and for which it does not.
The paper is organized as follows. In Sec.\ \ref{sec:II} we consider some
very general arguments to develop a criterion for the breakdown of LGW theory.
In particular, we discuss a relation between the analytic properties of
the order parameter susceptibility in the disordered phase and the 
applicability of LGW theory. In Sec.\ \ref{sec:III} we first formulate a
general field theory for interacting fermions that this criterion can
be applied to. We then consider three classes of zero-wavenumber order 
parameter fields, one each in the particle-hole
spin-singlet and spin-triplet channels, and one in the particle-particle
channel. We show that for the first class, LGW theory works, while for
the other two it does not.
In the last part of this section we show that these results are
consistent with explicit calculations.
In Sec.\ \ref{sec:IV} we discuss these results.
We discuss the generality of the model we consider, the coupling between
the order parameter fluctuations and other soft modes, and questions regarding
the coupling of statics and dynamics in quantum statistical mechanics. We
also discuss the effect of disorder on our results.
The paper closes with a series of final remarks. Perturbative
results for some susceptibilities relevant to our dicussion are 
summarized in Appendix \ref{app:A}.

\section{A Criterion for the Breakdown of Local LGW Theory}
\label{sec:II}

Of the two examples mentioned in the Introduction, the quantum ferromagnetic
and antiferromagnetic transitions, respectively, of itinerant electrons, the
ferromagnetic one has a zero-wavenumber, or homogeneous, order parameter,
while the antiferromagnetic one has a non-homogeneous order parameter. The
impossibility of constructing a local order parameter field theory for the
former is due to the coupling of fermionic soft modes to the order parameter
fluctuations, all of which are soft at the same (zero) wavenumber. For clarity
we stress that these fermionic modes are soft modes in addition to 
the critical order parameter fluctuations, and to any Goldstone modes in the
ordered phase. The coupling of these additional soft modes to the 
antiferromagnetic order parameter, which is an object at nonzero wavenumber, 
is much weaker and hence does not spoil the LGW concept in this case.

The above considerations imply that the existence of a local LGW theory will 
generically be much more questionable for quantum phase transitions with
homogeneous, or zero-wavenumber, order parameters, than for non-homogeneous
ones. In this paper we will therefore investigate the case
of quantum phase transitions with homogeneous order parameters.
The crucial question is then whether or not the coupling of the additional,
fermionic, soft modes to the order parameter fluctuations is strong enough 
to destroy the local nature of the LGW theory.

To sharpen this question, and put it in mathematical terms, 
we note that in Hertz's scheme the vertices of the
LGW theory are given by order parameter correlation functions in a reference
ensemble that consist of the full action minus an interaction term that has
been decoupled by a Hubbard-Stratonovich transformation (see 
Ref.\ \onlinecite{Hertz} and Sec.\ \ref{subsec:III.B} below). 
Since the interaction term one chooses for the decoupling is the one
that causes the phase transition, the reference system is always in the
disordered phase. The two-point, or ordinary, order parameter susceptibility
$\chi^{(2)}\equiv\chi$ in the reference ensemble determines the Gaussian 
vertex, and the higher order correlation functions $\chi^{(n)}$ ($n>2$) give 
the higher vertices. The locality, or otherwise, of the LGW theory then 
depends on the properties of the $\chi^{(n)}$. More precisely, since the
square-gradient term in the LGW functional comes from the wavenumber expansion 
of $\chi$, and the coefficients of the important (i.e., quadratic and cubic)
terms in the equation of state are determined by 
the zero-wavenumber and zero-frequency limits of $\chi^{(3)}$ and
$\chi^{(4)}$, it follows that in order for
the local LGW approach to break down, the wavevector dependent
susceptibility $\chi ({\bf q})$ must not
be an analytic function of $\vert{\bf q}\vert$ at ${\bf q}=0$, 
or $\chi^{(n)}$ ($n=3,4$) 
must diverge as ${\bf q}\rightarrow 0$. 
Such nonanalyticities can only arise from infrared
singularities, i.e., from fermionic soft modes. As has been discussed in 
Ref.\ \onlinecite{us_fermions},
these soft modes exist for various symmetry reasons, and if the symmetry
responsible for a particular soft mode is broken, the mode will acquire
a mass that depends on the symmetry-breaking parameter. For
reasons that will become clear below, 
let us suppose that the field $H$ conjugate to the order parameter
for the quantum phase transition, i.e., a source for the order parameter field,
breaks the above symmetry. Then the most general form of the
the singular part of $\chi$ that is consistent with power-law scaling is
form\cite{chi_sing_footnote,FMN_footnote}
\be
\chi_{\rm sing}({\bf q},H) = (\vert{\bf q}\vert^x + \vert H\vert)^y\quad. 
\label{eq:2.1}
\ee
Here $y,x>0$ are exponents that determine the nature of the singularity, and
the scaling of $H$ with the wavenumber, respectively.\cite{scaling_footnote} 
For example, for clean 
and disordered itinerant quantum ferromagnets in $d$ spatial dimensions, one 
has $x=1$, $y=d-1$, and $x=2$, $y=(d-2)/2$, respectively. (In the clean case 
in $3$-$d$, $y=2$ should be interpreted as 
$(\vert{\bf q}\vert + \vert H\vert)^2\ln (\vert{\bf q}\vert + \vert H\vert)$, 
see Eq.\ (\ref{eq:3.5}) and the discussion below it.) 
The higher susceptibilities $\chi^{(n)}$ can be obtained from $\chi$ by 
differentiating $(n-2)$ times with respect to $H$. This leads to infinite 
zero-field, zero-wavenumber $\chi^{(n)}$ for all $n>y+2$, and hence to a 
nonlocal field theory.

With these considerations we can now state and justify a criterion for
quantum phase transitions in itinerant electron systems that we 
will proceed to discuss and apply in the following sections of this paper.
\smallskip
\smallskip

\noindent
{\it Criterion}: Hertz theory (a local LGW theory) breaks down if
a quantum phase transition has a homogeneous, or zero-wavenumber, order 
parameter, and if a
source term $H$ for the order parameter field changes the soft-mode
spectrum of the fermionic or reference ensemble part of the action.
\smallskip
\smallskip

The validity of this criterion follows from our above discussion. We first
note that interacting electronic systems are intrinsically nonlinear, and
therefore generically all soft modes couple to all the physical quantities.
Therefore, if some soft modes are given a mass by $H>0$, i.e., are of 
the form $1/(\vert{\bf q}\vert^x + H)$, then the free energy and thus all
of the order parameter susceptibilities will be nonanalytic functions of $H$. 
This is only possible if the two-point order parameter susceptibility in the
reference ensemble has the form 
given in Eq.\ (\ref{eq:2.1}). It then follows that at $H=0$ all order 
parameter susceptibilities are singular functions of the wavenumber.
That is, a local LGW theory will not be possible.

Two remarks might be helpful at this point: (1)
The second condition in the above criterion ensures that there is
a sufficiently strong coupling between the fermionic soft modes and the 
order parameter fluctuations to invalidate the local LGW approach.
(2) Instead of referring to the soft-mode spectrum of the reference
ensemble, one could also demand that a nonzero $H$ changes the soft-mode
spectrum of the full action in the disordered phase away from the
critical point. Since the order parameter fluctuations, which are taken
out of the reference ensemble, are massive in this region, these two
requirements are equivalent.
 
To conclude this section we note that the analytical properties of various
susceptibilities in interacting electronic systems have been examined
in perturbation theory at $H=0$. The spin susceptibility
is known to be a nonanalytic function of the wavenumber at second order
in the screened interaction,\cite{us_chi_s} consistent with the known 
breakdown of Hertz theory for the itinerant quantum 
ferromagnet.\cite{us_fm_dirty,us_fm_clean} However,
the number density, number density current, and
number density stress susceptibilities have all been shown to not have
a nonanalytic wavenumber dependence to that order, see Appendix\ \ref{app:A}. 
This suggests that for quantum phase transitions with these observables as
order parameters, Hertz theory might work. While in principle it
is possible that nonanalyticities would appear in these correlation functions
at higher order in the perturbation theory, 
at least in the case of the number density current susceptibility this is very
unlikely, for reasons explained in Section \ref{subsec:III.D}. 
Also, in order to ruin the local LGW theory, any nonanalyticity 
would have to be cut off by the appropriate
external field. In the case of the number density, where the conjugate
field is just the chemical potential, this is hard to imagine.
In any case, the complexity of perturbation
theory makes it impractical to go beyond second order, and a more powerful
approach is needed to determine which quantum phase transitions can be
described by local order parameter field theories.

In the remainder of this paper we develop a classification scheme for 
quantum phase transitions with homogeneous order parameters that is based
on the above criterion. We will show that there are classes of quantum
phase transitions for which a local LGW functional exist, i.e., for which
Hertz theory is valid, and classes for which it breaks down.

\section{Classification Scheme for Quantum Phase Transitions with
         Homogeneous Order Parameters}
\label{sec:III}

\subsection{Fermionic Field Theory}
\label{subsec:III.A}

Our starting point is a general action for itinerant, interacting 
electrons,\cite{NegeleOrland}
\bml
\be
S = -\int dx\ {\bar\psi}(x)\left[\partial_{\tau} + \epsilon(\partial_{\bf x})
                                 - \mu\right]\,\psi(x)
    + S_{\rm int}\quad.
\label{eq:3.1a}
\ee
Here ${\bar\psi}(x)\equiv ({\bar\psi}_{\uparrow}(x),{\bar\psi}_{\downarrow}(x))$ 
and ${\psi}(x)\equiv (\psi_{\uparrow}(x),\psi_{\downarrow}(x))$ 
are fermionic (i.e., Grassmann-valued)
two-component spinor fields, and the index $x\equiv ({\bf x},\tau)$ 
comprises the real space position ${\bf x}$ and the imaginary time $\tau$.
$\int dx \equiv \int d{\bf x}\,\int_0^{\beta} d\tau$
with $\beta = 1/k_BT$, and the product of ${\bar\psi}$ and $\psi$ is understood
as a scalar product in spinor space that accomplishes the summation over the
two spin projections. $\mu$ is the chemical potential, and $\epsilon$
denotes the dispersion relation. For instance, for free electrons one has
$\epsilon(\partial_{\bf x}) = -\partial_{\bf x}^2/2m$ with $m$ the free
electron mass. 

$S_{\rm int}$ describes the electron-electron interaction,
which we will keep general. At the most basic level,
$S_{\rm int}$ is the Coulomb interaction, but often one starts at the level
of an effective theory, where some degrees of freedom have already been
integrated out to create short-ranged, effective interactions between
various modes. For our purposes, we only assume that $S_{\rm int}$ contains
an interaction between the order parameter modes. If we denote the order
parameter, in terms of the fermionic fields, by $n({\bar\psi},\psi)$,
this part of $S_{\rm int}$ reads schematically
\be
S_{\rm int}^{\rm OP} = J \int dx\ n^2(x)\quad,
\label{eq:3.1b}
\ee
with $J$ an appropriate coupling constant. Notice that in general $n$ will
be a tensor, so the notation $n^2$ in Eq.\ (\ref{eq:3.1b}) is symbolic.
For conceptual simplicity's sake, we also assume that the interacting
system in the disordered phase has a Fermi liquid ground state, i.e. that
the interactions are not sufficiently singular to destroy the Fermi liquid.
We will come back to this assumption in the Discussion, Sec.\ \ref{sec:IV}.

Finally, we write
\be
S = S_0 + S_{\rm int}^{\rm OP}\quad,
\label{eq:3.1c}
\ee
\eml
with $S_0$ containing all pieces of the action other than 
$S_{\rm int}^{\rm OP}$. $S_0$ describes the reference ensemble that
was alluded to in Sec.\ \ref{sec:II}. We note that, although the
order parameter interaction term is missing from the bare reference
ensemble action $S_0$, such an interaction is in general generated
in perturbation theory, albeit with a coupling constant that is
smaller than the critical value necessary for a nonvanishing expectation
value of the order parameter. The correlation functions of the reference
ensemble thus are those of the full system, with action $S$, in the
disordered phase.

\subsection{Order Parameter Field Theory}
\label{subsec:III.B}

Here we briefly review the derivation of the order parameter or LGW 
theory.\cite{Hertz} Let us decouple the order parameter interaction
$S_{\rm int}^{\rm OP}$, Eq.\ (\ref{eq:3.1b}), by means of a
Hubbard-Stratonovich field $M$. That is, we write the partition function
\bea
Z&=&\int D[{\bar\psi},\psi]\ e^{S[{\bar\psi},\psi]}
\nonumber\\
 &=&{\rm const.}\hskip -1pt\times\hskip -4pt\int\hskip -4pt 
     D[M]\ e^{-J\int dx\,M^2(x)}\left\langle 
     e^{- 2J\hskip -1pt \int\hskip -1pt dx M(x)\,n(x)}\right\rangle_{0},
\nonumber\\
 &\equiv&{\rm const.}\times\int D[M]\ e^{-\Phi[M]}\quad,
\label{eq:3.2}
\eea
where $\langle\ldots\rangle_0$ denotes an average with the reference ensemble
action $S_0$, and $\Phi[M]$ is the LGW functional. The latter reads explicitly
\be
\Phi[M] = J\int dx\ M^2(x) - \ln\left\langle e^{-2J\int dx\ M(x)n(x)}
          \right\rangle_0\quad,
\label{eq:3.3}
\ee
and can be expanded in powers of $M$,
\bml
\bea
\Phi[M]&=&\frac{1}{2}\int dx_1\,dx_2\ M(x_1)\biggl[\frac{1}{J}\,
          \delta(x_1-x_2)
\nonumber\\
&&\hskip 50pt - \chi^{(2)}(x_1-x_2)\biggr]\,M(x_2)
\nonumber\\
&&\hskip 5pt +\frac{1}{3\,!}\int dx_1\,dx_2\,dx_3\ \chi^{(3)}(x_1,x_2,x_3)
\nonumber\\
&&\hskip 24pt \times M(x_1)\,M(x_2)\,M(x_3) + O(M^4)\quad,
\label{eq:3.4a}
\eea
where we have scaled $M$ with $1/\sqrt{2}\,J$. The coefficients $\chi^{(l)}$
in the Landau expansion, Eq.\ (\ref{eq:3.4a}), are connected $l$-point 
correlation functions of $n(x)$ in the reference ensemble,
\be
\chi^{(l)}(x_1,\ldots,x_l) = \left\langle n(x_1)\cdots n(x_l)
                              \right\rangle_0^c\quad.
\label{eq:3.4b}
\ee
\eml

A crucial question now arises concerning the behavior of these correlation
functions in the limit of long distances and times or small frequencies and
wavenumbers. If their Fourier transforms are finite in that limit, then the
LGW functional $\Phi$ is local, and Hertz's analysis of the quantum phase
transition applies. However, this is not the case if the fermionic soft modes
that have been integrated out in the above procedure couple sufficiently 
strongly to the order parameter field. A prominent example is the case of a 
ferromagnetic order parameter, where $\chi^{(2)}$ is a nonanalytic function 
of the wavenumber,\cite{us_chi_s}
\be
\chi^{(2)}({\bf q}\rightarrow 0,\omega=0) \propto {\rm const.} 
   + \vert{\bf q}\vert^{d-1} + O({\bf q}^2)\quad.
\label{eq:3.5}
\ee
The integer exponents in $d=1$ and $d=3$ are to be interpreted as
$\ln(1/\vert{\bf q}\vert)$ and ${\bf q}^2\ln(1/\vert{\bf q}\vert)$,
respectively. Higher order correlation functions, starting with $\chi^{(4)}$, 
diverge in this limit like 
$\chi^{(n)} \sim \vert{\bf q}\vert^{-2(n-1)+d}$.\cite{us_fm_clean}
A nonzero expectation value $\langle M\rangle\neq 0$ in the ordered phase,
or an external field $H$ conjugate to $M$ in the disordered phase, cuts
off these nonanalyticities by giving a mass to some soft modes (to
spin-triplet particle-hole excitations in the ferromagnetic example). This 
leads to the free energy being a nonanalytic
function of $\langle M\rangle$ or $H$.\cite{us_fm_clean,us_1st_order} 
The above discussion is a more technical rerendering of the criterion
given in Sec.\ \ref{sec:II}. In what follows, we will use this criterion
as a diagnostic tool.

\subsection{Sources, and Symmetry Considerations}
\label{subsec:III.C}

We now add source or external field terms to our action, and ask whether
they change the soft mode structure of the system.
We will consider three explicit examples, corresponding to two classes of
homogeneous spin-singlet and spin-triplet order parameters in the
particle-hole channel, respectively, and a third class of particle-particle
channel order parameters.

\subsubsection{Particle-Hole Channel Spin-Singlet Order Parameters}
\label{subsubsec:III.C.1}

Consider a source term for a class of homogeneous, spin-singlet 
particle-hole order parameter fields,
\begin{equation}
S_H = H\int dx\ {\bar\psi}(x)\,f(\partial_{\bf x})\,\psi(x)\quad,
\label{eq:3.6}
\end{equation}
with $f$ an arbitrary polynomial function of the gradient operator.
Obviously, this source term has the same structure as the
most general dispersion term in the band electron part of the 
action $S_0$, the first term on the right-hand
side of Eq.\ (\ref{eq:3.1a}). Depending on the actual structure of
$\epsilon(\partial_{\bf x})$, $S_H$ may or may not break a 
spatial symmetry
of $S$. However, it does not change the soft-mode spectrum since any
electronic action with this structure describes a Fermi liquid, and
the soft modes of all Fermi liquids are in one-to-one correspondence
to one another. The free energy is therefore
an analytic function of $H$. It follows,
from the criterion in Sec.\ \ref{sec:II}, that for
any quantum phase transition with an order parameter of the form
\be
n(x) = {\bar\psi}(x)\,f(\partial_{\bf x})\,\psi(x)\quad,
\label{eq:3.7}
\ee
Hertz theory works, and the quantum critical behavior in $3$-$d$ is 
in general mean-field like. From the discussion in connection with
Eq.\ (\ref{eq:2.1}) it further follows 
that the order parameter susceptibility in the
disordered phase cannot have a nonanalytic wavenumber dependence that is
cut off by $H$.\cite{nonanalyticity_footnote}

An example of an order parameter in this class is the one for the 
isotropic-to-nematic phase transition that has been proposed to occur 
in quantum Hall systems by Oganesyan et al.\cite{Oganesyan_et_al}
For this phase transition Hertz theory works, and the mean-field critical 
behavior determined in Ref.\ \onlinecite{Oganesyan_et_al} is the exact quantum
critical behavior. Consistent with this, a perturbative calculation of 
the stress susceptibility $\chi_{xy}$ defined in Appendix \ref{app:A}, 
which plays the role of the reference ensemble susceptibility for the 
isotropic-to-nematic transition, found no nonanalytic wavenumber dependence.

\subsubsection{Particle-Hole Channel Spin-Triplet Order Parameters}
\label{subsubsec:III.C.2}

Now consider a class of source terms analogous to Eq.\ (\ref{eq:3.6}),
but in the spin-triplet channel,
\begin{equation}
S_{\vec H} = {\vec H}\cdot\int dx\ {\bar\psi}(x)\,f(\partial_{\bf x})\,
             {\vec\sigma}\, \psi(x)\quad,
\label{eq:3.8}
\end{equation}
with ${\vec\sigma} = (\sigma_x,\sigma_y,\sigma_z)$ the Pauli matrices.
Such terms break the invariance of $S$ under the SU(2) spin rotation group,
so the three components of the spin-triplet are no longer equivalent, and
transverse spin-triplet particle-hole excitations acquire a mass. For
instance, take ${\vec H}$ to point in $z$-direction, and consider
the temporal Fourier transform of $\psi(x)$, 
$\psi_n({\bf x})\equiv \psi({\bf x},\omega_n)$, with $\omega_n = 2\pi T(n+1/2)$
a fermionic Matsubara frequency. For ${\vec H}=0$, the homogeneous transverse
particle-hole susceptibility
\bea
&\int& d{\bf x}\,d{\bf y}\ \left\langle\left( 
     {\bar\psi}_n({\bf x})\,\sigma_x\,\psi_m({\bf x})\right)\left(
     {\bar\psi}_n({\bf y})\,\sigma_x\,\psi_m({\bf y})\right)\right\rangle\quad,
\nonumber\\
&&\hskip 160 pt (nm<0)\quad,
\nonumber
\eea
diverges for $\Omega_{n-m}\equiv\omega_n - \omega_m \rightarrow 0$ 
like $1/\Omega_{n-m}$. For a
nonvanishing source field, this soft mode acquires a mass proportional
to the magnitude of ${\vec H}$, i.e. the above susceptibility in the
zero-frequency limit has a finite value proportional to $1/\vert{\vec H}\vert$.

With an order parameter of the general form
\be
{\vec n}(x) = {\bar\psi}(x)\,f(\partial_{\bf x})\,{\vec\sigma}\,\psi(x)\quad,
\label{eq:3.9}
\ee
Hertz's LGW approach will therefore break down, the free energy is a 
nonanalytic function of the average order parameter, the quantum critical
behavior is in general not mean-field like, and the order parameter 
susceptibility in the disordered phase is a nonanalytic function of the
wavenumber.

The primary example for this class is the quantum itinerant ferromagnetic
transition that we have mentioned several times before. Other examples
include the spin-triplet analog of the isotropic-to-nematic transition
discussed in Ref.\ \onlinecite{Oganesyan_et_al}, for which Hertz theory
will not work, in contrast to the spin-singlet version.

\subsubsection{Particle-Particle Channel Order Parameters}
\label{subsubsec:III.C.3}

We finally consider a class of source terms for order parameters in the
particle-particle channel,
\be 
S_H = H \int dx\ \psi_{\sigma}(x)\,f(\partial_{\bf x})\,
                 \psi_{\sigma'}(x)\quad,
\label{eq:3.10}
\ee
which are relevant for superconductivity. These sources give a mass
to two-particle excitations in the particle-particle channel. 
For order parameters of the form
\be
n(x) = \psi_{\sigma}(x)\,f(\partial_{\bf x})\,\psi_{\sigma'}(x)\quad,
\label{eq:3.11}
\ee
Hertz theory will therefore break down. The prime example for this class
is the zero-temperature metal-to-superconductor transition.\cite{us_sc}

\subsection{Connection with Perturbative Results}
\label{subsec:III.D}

The above considerations imply that for the class of order parameters 
given by Eq.\ (\ref{eq:3.6}) one need not worry about a breakdown of 
Hertz theory, and the exact quantum critical behavior is easy to determine. 
In addition, they also explain a number of results concerning the presence 
or otherwise of nonanalytic wavenumber dependences in various 
susceptibilities, which show a pattern that was not understood before. 

The topic of a possible nonanalytic wavenumber and/or temperature dependence
of static correlation functions in a Fermi liquid has a long history, which
has been reviewed in Ref.\ \onlinecite{us_chi_s}. In this reference it was
also established that the wavenumber dependent spin susceptibility $\chi_s$ 
at zero temperature has the form given in Eq.\ (\ref{eq:3.5}), a result
that was confirmed by an explicit calculation in $2$-$d$ by 
Chitov and Millis.\cite{ChitovMillis} This was
done by means of perturbation theory to second order in the electron-electron
interaction. Reference \onlinecite{us_chi_s} also gave a physical argument, 
based on the coupling of
zero-sound modes, that links the ${\bf q}^2\ln\vert{\bf q}\vert$ dependence
in $3$-$d$, and the $\vert{\bf q}\vert^{d-1}$ dependence for general $d$,
to the $\ln\vert{\bf q}\vert$ dependence in $d=1$.\cite{DL} This
means that, unless a prefactor accidentally vanishes in some dimension, a 
correlation function that shows the former nonanalyticity in
$3$-$d$ will necessarily have the latter in $1$-$d$, 
as is the case for $\chi_s$.
Analogous calculations for various spin-singlet susceptibilities found no
nonanalyticity to second order in the 
interaction.\cite{us_chi_s,us_unpublished}. All of these results are summarized
in Appendix \ref{app:A}. Chitov and Millis\cite{ChitovMillis} speculated
that, at least in the case of the number density susceptibility, this null
result is an artifact of low-order perturbation theory, and that there actually
is a nonanalytic wavenumber dependence with the same strength as in the spin
density susceptibility, with the prefactor being of cubic or
higher order in the interaction. Given the computational effort of the
perturbation theory, this would be very hard to check explicitly. However,
in the case of the density current susceptibility there is a strong argument
against such a hypothesis: The f-sum rule, which reflects particle number
conservation, requires that the homogeneous density current susceptibility
is equal to $n_e/m$, with $n_e$ the electron density and $m$ the electron mass.
This means that for very fundamental reasons this susceptibility cannot have a 
$\ln\vert{\bf q}\vert$ singularity in $d=1$, and from the mode-mode coupling 
argument mentioned above it follows that therefore there cannot be a 
nonanalyticity stronger than $\vert{\bf q}\vert^x$ with $x>d-1$ in higher 
dimensions either.\cite{diamagnetism_footnote} 
Since the density and density
current susceptibilities are closely related by the same conservation law,
this casts serious doubt on a nonanalyticity in the former as well.

Our general arguments based on the soft-mode structure of the system
provide an explanation for all of these
results. In particular, they show that neither the density susceptibility,
nor any other spin-singlet susceptibility, to any order in perturbation
theory, has a nonanalytic wavenumber
dependence that is cut off by the appropriate conjugate field. While in
principle this leaves open the possibility of a nonanalyticity of a 
different nature,\cite{nonanalyticity_footnote} it makes it likely that
these susceptibilities are analytic at zero wavenumber, and the
perturbative results summarized in Appendix \ref{app:A} are consistent 
with this.

In the particle-particle channel, the susceptibility of the spin-singlet 
anomalous density has a nonanalytic wavenumber dependence that is cut off
by a superconducting gap, and accordingly
the metal-superconductor transition at zero temperature is not
described by Hertz theory.\cite{us_sc} 
This is again in agreement with the general
arguments given in Sec.\ \ref{subsubsec:III.C.3}. These calculations
for the particle-particle channel were for systems with quenched
disorder, but that does not affect our arguments, see Sec.\ \ref{subsec:IV.C}
below.

\section{Discussion}
\label{sec:IV}

\subsection{Generalizations}
\label{subsec:IV.A}

In this paper we have provided a general scheme to answer the question
of whether it is possible to construct a local LGW theory, i.e. a field 
theory solely in terms of the order parameter, for a given quantum phase
transition. We have applied this general philosophy to the one-band model
defined in Sec.\ \ref{subsec:III.A}, but more general models can easily be
analyzed in the same way.

As an example we consider a toy model with two completely degenerate bands
\bea
S &=& -\int dx\ \sum_{a=1,2} \bar\psi^{(a)}(x)\left[\partial_{\tau} 
      + \epsilon(\partial_{\bf x}) - \mu\right]\,\psi^{(a)}(x)
  \nonumber \\
&&+ \sum_{a=1,2} S_{\rm int}\left[\bar\psi^{(a)}, \psi^{(a)}\right]
\nonumber\\
&&    + S_{12}[\bar\psi^{(1)},\psi^{(1)},\bar\psi^{(2)},\psi^{(2)}]\quad.
\label{eq:4.1}
\eea
Here $a=1,2$ denotes the band index. In the simplest case,  
the inter-band interaction $S_{\rm 12}$ could be an interaction 
between the number or charge densities in the two bands
\be
S_{\rm 12} = J_{\rm 12} \int dx ~n_c^{(1)}(x) ~n_c^{(2)}(x) \quad,
\label{eq:4.2}
\ee
where $n_c^{(a)}(x)$ is the charge density in band $a$.
With increasing $J_{\rm 12}$ the system will undergo a quantum
phase transition from a symmetric state, in which the number densities in both 
bands are identical, to an asymmetric state with different densities in the 
two bands. The order parameter for the transition is the density difference
$n_c^{(1)}(x) - n_c^{(2)}(x)$.

Let us now apply the soft-mode considerations developed in Sec. \ref{sec:III} to
this model. A source term for the order parameter field,
\be
S_{H_{12}} = H_{12} \int dx\ [n_c^{(1)}(x) - n_c^{(2)}(x)]\quad,
\label{eq:4.3}
\ee
breaks the symmetry between the two bands manifest in
(\ref{eq:4.1}) and changes the soft
mode structure. Therefore, the band symmetry breaking quantum phase transition
will not be described by a local order parameter field theory.

We also note that this transition is very similar to the magnetic transitions
discussed above. If one neglects the spin degrees of freedom in the two-band
model, it can be mapped onto a quantum ferromagnetic transition
with Ising symmetry by identifying the band indices 1 and 2 with spin-up and 
spin-down, respectively.

\subsection{Coupling between Soft Modes and Order Parameter Fluctuations}
\label{subsec:IV.B}

The mechanism for a breakdown of the LGW approach to quantum phase 
transitions that we have studied is soft modes that are
integrated out in deriving an LGW theory. It is important to realize that
such soft modes {\em always} exist in itinerant electron systems, 
and in general they always couple to
the order parameter fluctuations via mode-mode coupling effects, except
for special cases where such a coupling is forbidden by some symmetry.
The crucial question is whether this coupling is strong enough to lead
to the free energy being a nonanalytic function of the order parameter.
The criterion for this is whether a nonzero average value of the order
parameter, or, equivalently, a nonzero external field conjugate to the
order parameter, gives a mass to the soft modes in question. This is more
easily accomplished if both the order parameter and the additional soft
modes are massless at the same wavenumber, as is the case for the
itinerant quantum ferromagnet. A counterexample is the quantum antiferromagnet,
where the order parameter is the staggered magnetization, a finite-wavenumber
quantity that couples only weakly to the soft particle-hole excitations that
are soft at zero wavenumber.

\subsection{Effects of Disorder}
\label{subsec:IV.C}

So far we have discussed clean systems, but all of our methods 
remain valid in the presence of quenched disorder, and so does our
discussion. The only modification is that in the presence of quenched
disorder the Fermi liquid ground state is destroyed for dimensions
$d\leq 2$ rather than $d\leq 1$. Quenched disorder is described
by a term in the action
\be
S_{\rm dis} = \int dx\ u({\bf x})\,{\bar\psi}(x)\,\psi(x)\quad,
\label{eq:4.4}
\ee
with $u({\bf x})$ a random potential. The source term for a homogeneous
spin-singlet order parameter, Eq.\ (\ref{eq:3.6}), still does not change the
soft-mode structure of the disordered system (although we stress that the
latter is different from that of a clean system), and consequently
the quantum phase transitions with such order parameters in disordered
systems can be described by local LGW theories. Furthermore, the corresponding
susceptibilities are expected to be analytic functions of the wavenumber.
This is consistent with the fact that the static density susceptibility has no
$\vert{\bf q}\vert^{d-2}$ nonanalyticity in perturbation theory, 
and its homogeneous limit,
$\partial n/\partial\mu$, is finite in $2$-$d$.\cite{AA,F,R} Similarly, 
the spin-triplet source, Eq.\ (\ref{eq:3.8}), still breaks spin rotation
invariance and changes the soft-mode structure by giving the transverse
particle-hole excitations in the spin-triplet channel a mass. Consequently, 
LGW theory breaks down for the disordered itinerant
quantum ferromagnetic transition, and the quantum critical behavior is
not mean-field like. Consistent with this, in perturbation theory 
the spin susceptibility 
has a $\vert{\bf q}\vert^{d-2}$ nonanalyticity for $d>2$, and a 
$\ln\vert{\bf q}\vert$ behavior in $2$-$d$.\cite{AA,us_fm_dirty}

\subsection{Statics versus Dynamics}
\label{subsec:IV.D}

It is well known that in quantum statistical mechanics, statics and dynamics
are coupled. Naively, this might lead to the expectation that correlation
funtions that do not show a nonanalytic wavenumber dependence at zero
frequency have no nonanalytic frequency dependence either. This, however,
is not true. In quenched disordered systems, the real part of the 
electrical conductivity, which is the imaginary part of the dynamical
density current susceptibility divided by the frequency, has an
$\omega^{(d-2)/2}$ frequency dependence, even though the static
current susceptibility, as pointed out above, has no analogous
wavenumber nonanalyticity. Similary, the dynamical counterpart of the stress
susceptibility $\chi_{xy}$ describes the sound attenuation coefficient,
and it also is a nonanalytic function of the frequency.\cite{sound_attenuation} 
It is therefore
{\em not} correct to conclude from the presence of a nonanalytic frequency
dependence in a time correlation function that the corresponding static
susceptibility will be a nonanalyic function of the wavenumber, and that
hence for the phase transition with the corresponding order parameter LGW
theory will break down.

In order to understand this asymmetry between wavenumber and frequency
dependences it is important to remember
that, even in quantum statistical mechanics, statics and dynamics are not
equivalent. It is known that a finite frequency in the fermionic field
theory breaks the symmetry between retarded and advanced Green 
functions, and gives a mass to soft modes.\cite{us_fermions} 
Therefore, considering
a {\em dynamical} source, as opposed to the static sources in
Eqs.\ (\ref{eq:3.6}) and (\ref{eq:3.8}), changes the situation. 
If one expands a dynamical spin-singlet source in powers of
$\omega$, then the zeroth order term does not change the soft-mode spectrum, 
but the term linear in $\omega$ does, and hence
the dynamical piece of a spin-singlet susceptibility has in general 
nonanalyticities, even though the static part does not. 
The electrical conductivity and the sound attenuation
coefficient mentioned above are examples of this effect.

\subsection{Final Remarks}
\label{subsec:IV.E}

We conclude with a few final remarks. First, we have restricted ourselves
to systems where the ground state in the disordered phase is a Fermi liquid,
but we have not explicitly used this property. We expect our general
method to still work in more general cases. As an example, consider the
case of a metamagnetic transition, i.e. a magnetic transition inside a
ferromagnetic phase. In this case the spin rotation symmetry is already
broken on either side of the transition, and adding the source term,
Eq.\ (\ref{eq:3.8}), does {\rm not} further change the soft-mode spectrum.
We therefore conclude that in this case LGW theory will work, in agreement
with the recent treatment of this situation by Millis et al.\cite{Millis_et_al}

Second, we have restricted our explicit discussion 
to homogeneous order parameters. As we have mentioned in the context
of the antiferromagnetic phase transition, inhomogeneous order parameters
are in general expected to couple less strongly to the fermionic soft
modes than homogeneous ones. 
However, the consequences of a spontaneous breaking of the 
translational invariance
warrant a more thorough investigation. This is of interest, for instance,
for the nature of the transition to the stripe phases that have been
predicted and observed to occur in high-temperature 
superconductors.\cite{Emery_et_al} These questions can
also be analyzed within
the framework set up in this paper.

\acknowledgments

This work was supported by the NSF under grant numbers DMR-98-70597 and 
DMR-99-75259, and by the DFG under grant number Vo 659/3. This
work was initiated at the Aspen Center for Physics, and we thank the
Center for its hospitality.

\appendix
\section{Perturbative results for order parameter susceptibilities}
\label{app:A}

In this appendix we list perturbative results for various susceptibilities.
In Ref.\ \onlinecite{us_chi_s} it was shown that the static spin 
density susceptibility $\chi_s$ has a nonanalytic wavenumber dependence 
in $3$-$d$,
\be
\chi_s({\bf q}) = 2N_F\,\left[1 + c_s({\bf q}/2k_F)^2\ln(2k_F/\vert{\bf q}
                  \vert) + O({\bf q}^2)\right]\quad,
\label{eq:A.1}
\ee
with $N_F$ the density of states per spin at the Fermi surface, $k_F$ the
Fermi wavenumber, and $c_s$ a positive constant that, for weak 
interactions, is quadratic
in the interaction amplitude. This result was confirmed in 
Ref.\ \onlinecite{ChitovMillis}, which also showed that in $2$-$d$ there is
a corresponding $\vert{\bf q}\vert$-nonanalyticity. All of these results
are consistent with a mode-mode coupling argument that lets one expect
a singularity of the the form $\vert{\bf q}\vert^{d-1}$, with the integer
exponents in $d=1$ and $d=3$ corresponding to logarithms,
see Eq.\ (\ref{eq:3.5}). In particular,
this argument links the perturbative $\ln\vert{\bf q}\vert$
dependence in $d=1$ (see Ref.\ \onlinecite{DL}) 
to the more general nonanalyticity in any dimension.

The perturbation theory developed in Ref.\ \onlinecite{us_chi_s} is
readily generalized to calculate other susceptibilities. For the number
density susceptibility $\chi_n$ one finds that, to quadratic order in the
interaction, the terms of order ${\bf q}^2\ln\vert{\bf q}\vert$ 
cancel,\cite{us_chi_s} leaving one with the behavior
\be
\chi_n({\bf q}) = 2N_F\,\left[1 + O({\bf q}^2)\right]\quad.
\label{eq:A.2}
\ee

We have also calculated the number density current susceptibility $\chi_j$
and the stress susceptibility $\chi_{xy}$ 
that is the static correlation function of
the electronic stress operator
$$
(2/k_F^2)\sum_{\bf k}k_xk_y\,c^{\dagger}_{{\bf k}+{\bf q}/2}\,
        c_{{\bf k}-{\bf q}/2}\quad.
$$
For both of these 
we have found that the same cancellations hold as in the case of the
number density susceptibility. That is, to second order in the interaction,
\bea
\chi_j({\bf q})&=&(n_e/m)\,\left[1 + O({\bf q}^2)\right]\quad,
\label{eq:A.3}\\
\chi_{xy}({\bf q})&=&(8N_F/15)\,\left[1 + O({\bf q}^2)\right]\quad,
\label{eq:A.4}
\eea
with $n_e$ the electron density and $m$ the electron mass.
The null results expressed by Eqs.\ (\ref{eq:A.2}) - (\ref{eq:A.4})
are valid for interaction amplitudes with arbitrary wavenumber and
frequency dependences.

In the particle-particle channel, the susceptibility of the
anomalous density $\psi_{\uparrow}(x)\,\psi_{\downarrow}(x)$ is
known to have a leading wavenumber dependence proportional to
$1/\ln\vert{\bf q}\vert$,\cite{us_sc} but no exact perturbative
results are available.

\end{document}